\documentstyle[prl,aps,twocolumn,epsfig]{revtex}  

\tighten       

\begin{document}     
\draft 
     
\title{Breaking the symmetry in 
bimodal frequency distributions of globally coupled oscillators
}     
\author{ J.\ A.\ Acebr\'on  and L.\ L.\ Bonilla }     
\address{Escuela Polit\'ecnica Superior,  Universidad Carlos III de
Madrid,     
Butarque 15, \\ 28911 Legan{\'e}s, Spain} 
\author{ S.\ De Leo }     
\address{Dipartimento di Fisica, Universit\`{a} di Lecce and INFN, Via
Provinciale Lecce-Arnesano, \\ 73100 Lecce, Italy}    

\author{ R.\ Spigler 
\cite{spigler:email}}     
\address{Dipartimento di Matematica, Universit\`{a} di ``Roma Tre",
Largo S. Leonardo Murialdo,1. \\ 00146 Roma, Italy} 
\narrowtext
\date{\today}     
\maketitle     
\begin{abstract} 

The mean-field Kuramoto model for synchronization of 
phase oscillators with an {\it asymmetric} bimodal frequency 
distribution is analyzed. Breaking the reflection symmetry 
facilitates oscillator synchronization to rotating wave phases. 
Numerical simulations support the results based of bifurcation 
theory and high-frequency calculations. In the latter case, 
the order parameter is a linear superposition of parameters 
corresponding to rotating and counterrotating phases.

\end{abstract}     
     
\pacs{ 05.45.+b, 05.20.-y, 64.60.Ht}               
\narrowtext
Collective synchronization and incoherence in large populations of 
nonlinearly coupled oscillators received a great attention in the 
recent years. Motivation for this can be found in the broad variety of
phenomena which can be modeled in this framework. Indeed, 
synchronous flashing in swarms of fireflies \cite{buck}, crickets that
chirp in unison \cite{walker}, epilectic seizures in the brain
\cite{traub},
electrical synchrony among cardiac pacemaker cells \cite{michaels}, 
arrays of Josephson junctions \cite{wiesen}, 
chemical processes \cite{scheu}, some models of charge density waves in 
quasi-one-dimensional metals \cite{bon1}, and some neural 
networks used to model dynamic learning processes\cite{arenas}, all 
seem to be described in these terms.

The mathematical model conceived first as a large collection 
of elementary nonlinear phase oscillators, each with a 
globally attracting limit-cycle, goes back to Winfree \cite{winfree}. 
It was later formulated as a system of nonlinearly coupled differential 
equations by Kuramoto \cite{kuramoto}, in the 
mean-field coupling case, and as a system of Langevin equations, 
(adding external white noise sources), by Sakaguchi \cite{sakaguchi},
\begin{equation}     
  \dot{\theta_{i}} = \omega_{i} + \xi_{i}(t)       
   + \frac{K}{N} \sum_{j=1}^{N}  \sin(\theta_{j} -\theta_{i}),
\quad   i=1,\ldots,N.  \label{1}      
\end{equation}  
Here, $\theta_{i}(t)$ denotes the {\it i}th oscillator phase,
$\omega_{i}$ its natural frequency (picked up from a given 
distribution $g(\omega)$), $K>0$ represents the coupling strength, 
and the $\xi_{i}$'s are independent identically distributed white 
noises. Consequently, the one-phase oscillator probability density, 
$\rho(\theta,t,\omega)$, obeys the following nonlinear
Fokker-Planck equation, in the thermodynamic limit $N \to \infty$: 
\begin{equation}     
       \frac{\partial \rho}{\partial t}=D \frac{\partial^{2}
\rho}{\partial      
\theta^{2}} -		\frac{\partial}{\partial \theta}(v \rho),
\label{2}     
\end{equation}
where $D>0$ comes from the noise terms in (\ref{1}), and
\begin{equation}
v(\theta,t,\omega)=\omega +K r \sin(\psi-\theta).
\end{equation}
Here the complex-valued order parameter, $r e^{i \psi}$, is defined by
\begin{equation}
r e^{i \psi}=\int_{0}^{2\pi} \int_{-\infty}^{+\infty} e^{i \theta}
\rho(\theta,t,\omega)  g(\omega) d\omega d\theta.\label{order}
\end{equation}
It is understood that (\ref{2}) must be accompanied by the prescription
of the initial value $\rho(\theta,0,\omega)=\rho_{0}(\theta,\omega)$, 
$2\pi$-periodic boundary conditions, and normalization $\int_{0}^{2\pi}
\rho(\theta,t,\omega) d\theta=1$. 

The fundamental phenomenon of transition from incoherence [$\rho 
\equiv 1/(2\pi)$, $r\equiv 0$] to collective synchronization 
($r\neq 0$) is similar to phase transitions in Statistical Physics, 
and it was first analyzed rigorously by Strogatz 
and Mirollo \cite{strog}. They studied the linear stability of
incoherence of populations characterized by unimodal frequency
distributions. In \cite{bns}, a nonlinear stability analysis was 
accomplished, and bimodal frequency distributions [$g(\omega)$ 
with two peaks] were also considered. In the latter case, new 
bifurcations were discovered, showing the existence of a rich 
phenomenology, such as subcritical spontaneous stationary 
synchronization, supercritical time-periodic synchronization, 
bistability and hysteretic phenomena. A large amount of information 
was obtained in \cite{strog,bns}, adopting as models of uni- and 
bi-modal frequency distributions, $g(\omega)=\delta(\omega)$, 
and $g(\omega)=\frac{1}{2}[\delta(\omega+\omega_{0})+ 
\delta(\omega-\omega_{0})]$. It may be surprising now to realize 
that the {\it asymmetric bimodal} distribution,
\begin{equation}
g(\omega)=\alpha \delta(\omega-\omega_{0}) +(1-\alpha)
\delta(\omega+\omega_{0}),
\label{asimet}
\end{equation}
entails essentially different features with respect to the symmetric 
case, $\alpha=\frac{1}{2}$, even for $\alpha$ close to $\frac{1}{2}$. 
Since, due to unavoidable imperfections, possibly small deviations 
from symmetry are most likely in Nature, the {\it asymmetric} case
should be rather ubiquitous.
The purpose of this paper is to illustrate the distinctive features of
an asymmetric oscillator frequency distribution. The main qualitative 
effect of asymmetry is that no synchronized stationary phase is 
possible. Synchronized phases branch off from incoherence as traveling
waves (TW, see below) and their structure becomes richer as the strength
of the coupling increases. Asymmetry of the frequency distribution 
changes the stability boundaries of the incoherence (see the phase 
diagrams in Figs.\ 1 and 2), rendering it less stable, and,
consequently, 
rendering the partially synchonized solution (whose order parameter, 
however, now depends always on time) more stable. 

The stability boundaries
for the incoherent solution, $\rho_0(\theta,\omega)\equiv 1/(2\pi)$ 
can be calculated by setting to
zero the greatest of the $Re (\lambda)$'s, where $\rho = \rho_0 +
\epsilon\, e^{\lambda t}\, \eta(\theta,\omega)$ ($\epsilon\to 0$), and
$\lambda$ are the eigenvalues of the
linearized problem. They are given by~\cite{strog}
\begin{equation}
\frac{x}{2} \int_{-\infty}^{+\infty} d\omega \frac{g(\omega)}{Z+i
\frac{\omega}{D}}=1,
\label{stab}
\end{equation}
where we have defined $x=K/D$ and $Z = 1+ \lambda/D$.
For the {\it asymmetric} bimodal distribution (\ref{asimet}), we find
\begin{equation}
Z_{1,2}=\frac{x}{4} \pm \left[\frac{x^{2}}{16}-y^{2}+ i \frac{x y}{2}
(2\alpha-1)\right]^{1/2}\,,
\end{equation}
where $y=\omega_0/D$. The stability regions in figs.\ 1, 2 are then
determined 
by the condition $\quad$ max $Re \ Z_{1,2}\ \leq 1$. 
\begin{figure}
\centerline{\hbox{\psfig{file=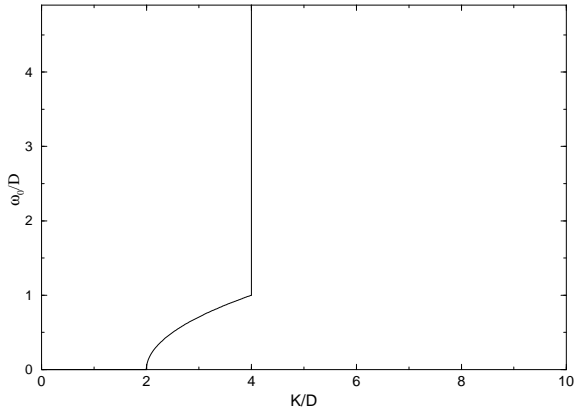,width=8.0cm}}}
\caption{Stability boundaries for the incoherent
solution for the {\it symmetric} bimodal frequency distribution.
Incoherence is
linearly stable in the region to the left of the solid line.}
\label{fig1}
\end{figure}\ 
\begin{figure}
\centerline{\hbox{\psfig{figure=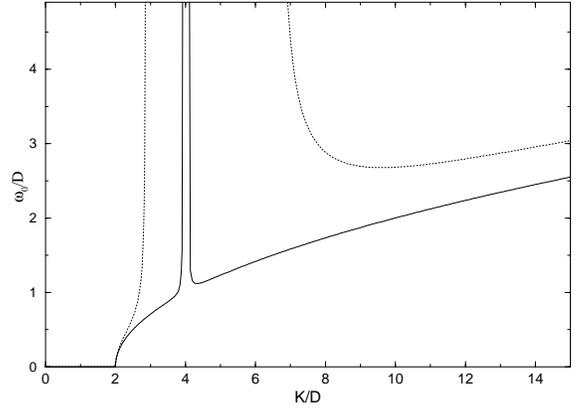,width=8.0cm}}}
\caption{Stability boundaries for the incoherent for the {\it
asymmetric} bimodal frequency distribution, $\alpha=0.49$ (solid 
line), $\alpha=0.3$ (dotted line).}
\label{fig2}
\end{figure}\ 

The branch on the right of the asymptote in fig. 2  is not 
completely unexpected. Indeed, its counterpart in the
symmetric case is a parabolic profile continuing that in fig.1
(see \cite{bns}). In the latter case, however, such a branch is 
not as important as in fig. 2, since it does not separate different 
stability regions. The behavior depicted in figs.\ 1 and 2 is confirmed
by direct numerical simulation \cite{acebron,ritort} of the Kuramoto-Sakaguchi 
equation (\ref{2}); see the evolution of the amplitude and phase of the order 
parameter in figs.\ 3, 4, and 5.

\begin{figure}
\centerline{\hbox{\psfig{figure=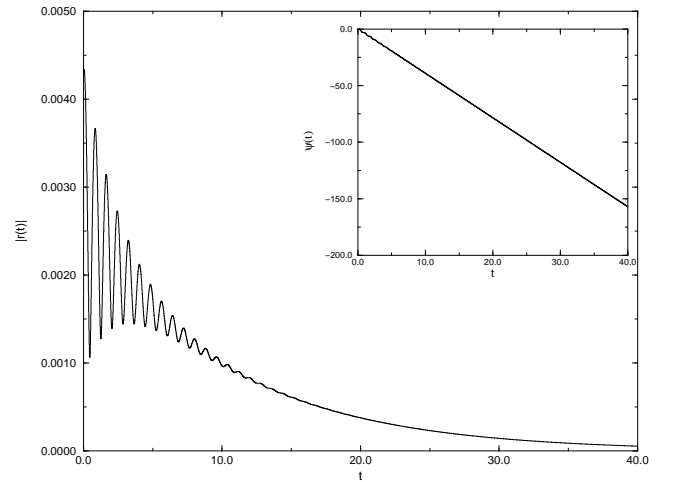,width=8.0cm}}}
\caption{Time evolution of the amplitude $|r(t)|$ and phase $\psi(t)$ of
the order parameter for 
$K=3,D=1,\omega_0=4$, with $\alpha=0.4$. Note the stability of the
incoherent solution.}
\label{fig3}
\end{figure}\ 
\begin{figure}
\centerline{\hbox{\psfig{figure=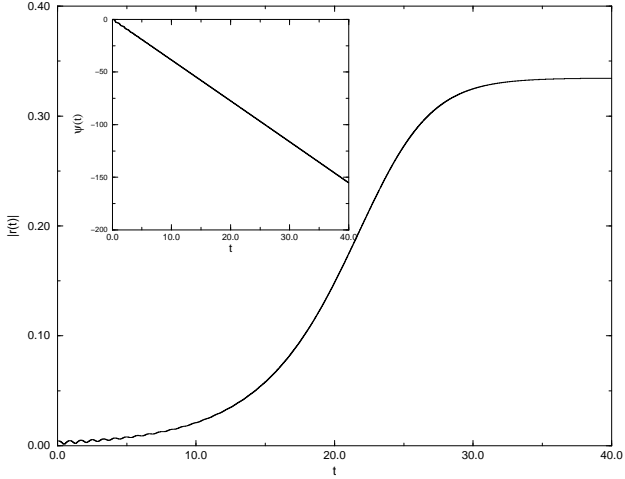,width=8.0cm}}}
\caption{Time evolution of the amplitude $|r(t)|$ and phase $\psi(t)$ of
the order parameter for 
$K=4,D=1,\omega_0=4$, with $\alpha=0.4$. This is the region of stability
of the TW solution}
\label{fig4}
\end{figure}\ 
\begin{figure}
\centerline{\hbox{\psfig{figure=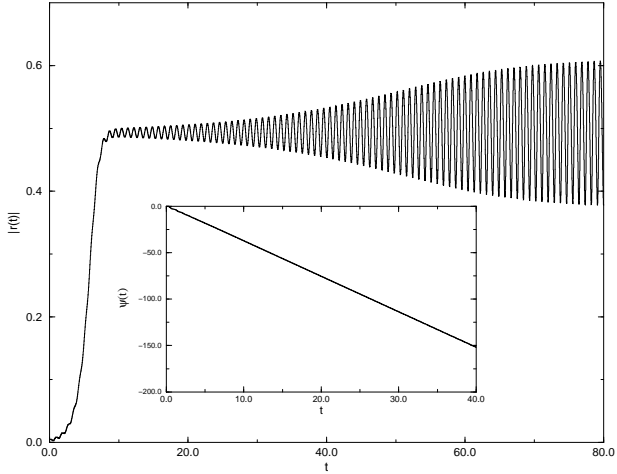,width=8.0cm}}}
\caption{Time evolution of the amplitude $|r(t)|$ and phase $\psi(t)$ of
the order parameter for 
$K=6,D=1,\omega_0=4$, with $\alpha=0.4$. Another TW solution has 
bifurcated from incoherence resulting in a more complex behavior of the
order parameter.}
\label{fig5}
\end{figure}\ 
Observe that the phase $\psi(t)$ is always time-dependent, rather than a
constant
as in the case of the symmetric bimodal distribution~\cite{bns,bps}. The
new 
synchronized phases are described by a bifurcation
analysis near the line in the parameter space where the incoherence
loses stability:
\begin{equation}
\frac{\omega_0}{D}=\frac{2(1-\frac{K_c}{4D})\sqrt{\frac{K_c}{2D}-1}}{
\sqrt{\alpha(1-\alpha)}\sqrt{(\frac{2}{\alpha}-\frac{K}{D})
(\frac{2}{1-\alpha}-\frac{K}{D})}}.\, \label{bif.line}
\end{equation}

This is obtained when the largest real part of the eigenvalues is set to
zero, and corresponds to $K_c=4D$, $\omega_0>D$, of the symmetric case 
(see fig.\ 1). The two-time asymptotic analysis conducted in \cite{bns}
may be used unchanged for bifurcations at the line (\ref{bif.line}) with 
the asymmetric frequency distribution, taking into account that now
$\Omega^2=\omega^2 + D^2 - K_c D/2$,
and that $g(\omega)$ is the asymmetric frequency distribution in
(\ref{asimet}). In fact, the symmetric case possesses the reflection 
symmetry $\omega_0\to - \omega_0$, $\theta\to - \theta$, which causes 
the eigenvalues to be doubly degenerated~\cite{craw}, whereas this is
{\it not} 
the case for the asymmetric frequency distribution. Then, the simple 
analysis of Ref.~\cite{bns} (which overlooked eigenvalue multiplicity, 
as pointed out in \cite{craw}; see also \cite{bps})
can be directly used for the asymmetric case. The result is that a 
branch of stable synchronized phases bifurcates from incoherence at the
point given by 
(\ref{bif.line}). Near the bifurcation line, these solutions have the
form of TWs rotating counterclockwisely~\cite{bns}:
\begin{eqnarray}
\rho(\theta,t,\omega) = \frac{1}{2\pi} + \frac{ R \, e^{i\Psi_{0} 
(K-K_{c}) (t-t_{0})}}{2\pi[D+i(\omega+\Omega)]} 
e^{i(\Omega t +\theta)} + c.c. \label{TW}\\
+ O(K-K_c),\nonumber
\end{eqnarray}
where c.c. means taking complex conjugate of the preceding term, and 
\begin{eqnarray}
R =\sqrt{\frac{(K-K_{c})\,\mbox{Re}\lambda_1}{\mbox{Re}\gamma}} ,
\quad \Psi_0 = \mbox{Im} \lambda_1 - \mbox{Im} \gamma \frac{
\mbox{Re}\lambda_1}{\mbox{Re} \gamma} \,. 
\label{par}
\end{eqnarray}
See Ref.\ \cite{bns} for the explicit expressions of the parameters
$\gamma$ and $\lambda_1 = (\partial\lambda/\partial K)|_{K=K_{c}}$. 
In the symmetric 
case, another solution corresponding to waves 
rotating clockwisely has to be added to (\ref{TW}). This results
in a stable standing wave solution, whose order parameter has a 
constant phase and an oscillatory amplitude~\cite{craw,bps}. 

In the high-frequency limit, $\omega_0 \rightarrow\infty$, a different 
perturbation analysis provides expressions for the evolution of the 
probability density, either near or far from bifurcation
points~\cite{acebron}. 
The main result is that the frequency distribution decomposes in 
as many phases as peaks of the oscillator frequency distribution in 
such a limit. Each phase rotates with the frequency corresponding
to its respective peak. Then, the order parameter may be written as a
linear 
superposition of the order parameters of the different phases. For the 
asymmetric bimodal distribution, the overall order parameter evolves 
(except by a constant phase shift) to 
\begin{equation}
r\, e^{i\psi} = \alpha R_+ e^{i\omega_{0} t +\Psi_{+}} + 
(1-\alpha) R_- e^{- i\omega_{0} t +\Psi_{-}},\label{linearsup}
\end{equation}
where $R_{\pm}$ and $\Psi_{\pm}$ correspond to phases rotating with
angular 
speeds $\pm\omega_0$. They can be calculated with the stationary formule
(2.1) and (1.7) of Ref.~\cite{bns}, with zero frequency~\cite{acebron}. 
Let $\alpha < 1/2$ to be specific. We have the 
following possibilities depending on the value of the coupling constant:
\begin{enumerate} 
\item If $0<K< 2D/(1-\alpha)$, the incoherent solution $\rho_0 \equiv
1/(2\pi)$
is stable and it is the only possible stationary solution. 
\item If $2D/(1-\alpha) < K < 2D/\alpha$, a globally stable partially  
synchronized solution branches off incoherence at $K=2D/(1-\alpha)$. 
It has $R_+ = 0$, $\psi = \Psi_- - \omega_0 t$, and $r = (1-\alpha)\,
R_-$. 
Its component $\rho_+ \equiv 1/(2\pi)$ is incoherent, while its
component 
$\rho_-$ is synchronized. The overall 
effect is having a TW solution (rotating clockwisely). 
\item If $K> 2D/\alpha$, the component $\rho_+$ becomes partially 
synchronized too. The probability density then has TW
components rotating clockwisely and anticlockwisely. Their order 
parameters have different strengths, and $R_- >R_+$ if $\alpha<1/2$. 
\end{enumerate}

Let us now compare the analytical results obtained in the high-frequency 
limit with those obtained by means of bifurcation theory. As 
$\omega_0\to\infty$, the parameters $\lambda_1,\gamma$, and
$K_c$ in  (\ref{par}) become (cf.\cite{bns}),
\begin{eqnarray}
\lambda_1=\frac{2D^2}{(1-\alpha)K_c^2}\,, \quad \gamma=\frac{1}{2D}\,,
\quad K_c=\frac{2D}{1-\alpha}. \nonumber
\end{eqnarray}
We can now calculate the order parameter in (\ref{order}) by using
(\ref{TW}), (\ref{par}), and the previous expression: 
\begin{eqnarray}
r e^{i \psi} 
\approx (1-\alpha)\,\left\{{K\, (1-\alpha)\over D} - 2\right\}^{{1\over
2}}
\, e^{-i \omega_{0} t}.
\label{comp}\end{eqnarray}
Eq.\ (\ref{comp}) agrees exactly with the results of the high-frequency
limit~(\ref{linearsup}) in \cite{acebron}: The
amplitude of the order parameter
is constant, and its phase decreases linearly in time. Of course, for
larger values of the coupling constant another branch of oscillatory
solutions (TW rotating clockwisely) bifurcates from incoherence.
Then, the overall probability density is richer, with an order parameter
whose amplitude and phase both vary with time as in Fig.\ 5.

In conclusion, we have analyzed the mean-field Kuramoto-Sakaguchi model
of
oscillator synchronization with an {\it asymmetric bimodal} frequency 
distribution. In this case, reflection symmetry is broken, which 
results in stable synchronized phases that have the form of TWs 
(rotating clockwisely or anticlockwisely). These waves have order
parameters
with constant amplitude, and phases which depend linearly on time. 
As the strength of the coupling constant increases, such a synchronized 
phase bifurcates from incoherence. Larger values of the coupling
strength 
result in a new bifurcation, which contributes to another TW. Then,  
both phase and amplitude of the order parameter become time-dependent. 
Numerical simulations of the model favorably agree with the results of
bifurcation theory, and of high-frequency perturbation expansions. 
Extensions of our analyses to the case of a multimodal frequency 
distribution [i.e., a discrete, or a continuous $g(\omega)$ having $m$ 
peaks] are worth considering in future works.
 
This work was supported, in part by the Italian GNFM-CNR (J.A.A. and
R.S.),
 the Fundaci\'on Carlos III de Madrid (J.A.A.), the Spanish
DGES under grant PB95-0296 (L.L.B.), and the EC Human Capital 
and Mobility Programme under contract ERBCHRXCT930413 (S.D.L.).
J.A.A. is grateful to University of ``Roma Tre", Rome, Italy, for its
hospitality
while Visiting Junior researcher of GNFM-CNR.


\begin{references} 
     
\bibitem[*]{spigler:email} {E-address \tt spigler@ulam.dmsa.unipd.it}.
Author to     
whom all correspondence should be addressed. 
\bibitem{buck} J.Buck, Quart. Rev. Biol. {\bf 63}:265 (1988).

\bibitem{walker} T.J. Walker, Science {\bf 166}:891 (1969).

\bibitem{traub} R.D. Traub, R. Miles, and R.K.S Wong, Science {\bf
243}:1319 (1989).

\bibitem{michaels} D.C. Michaels, E.P. Matyas, and J.Jalife, Circulation
Res. {\bf 61}:704 (1987).


\bibitem{wiesen} K. Wiesenfeld, P. Colet, and S.H. Strogatz, Phys. Rev.
Lett. {\bf 76}:404 (1996).

\bibitem{scheu} M. Scheutzow, Prob. Theory Related Fields, {\bf 72}:425
(1986).

\bibitem{bon1} L.L. Bonilla, Phys. Rev. B {\bf 35}:3637 (1987).

\bibitem{arenas} C.J. P\'erez Vicente, A. Arenas, and L.L. Bonilla, J.
Phys. A: Math. Gen. {\bf 29}:L9 (1996).

\bibitem{winfree} A.T. Winfree, Geometry of Biological Time,
Springer-Verlag, New York (1990).

\bibitem{kuramoto} Y. Kuramoto, in International Symposium of
Mathematical Problems in Theoretical Physics, H. Araki ed.,
Lecture Notes in Physics, Vol.39 (Springer, New York), (1975).

\bibitem{sakaguchi} H. Sakaguchi, Prog. Theor. Phys. {\bf 79}:39 (1988).



\bibitem{strog} S.H. Strogatz and R.E. Mirollo, J. Stat. Phys. {\bf
63}:613 (1991).

\bibitem{bns} L.L. Bonilla, J.C. Neu, and R. Spigler, J. Stat. Phys.
{\bf 67}:313 (1992).

\bibitem{craw} J.D. Crawford, J. Stat. Phys. {\bf 74}:1047 (1994).

\bibitem{bps} L.L. Bonilla, C.J. P\'erez Vicente, and R. Spigler,
Physica D (1996) submitted.


\bibitem{acebron} J.A. Acebr\'on and L.L. Bonilla, Physica D (1997)
submitted.

\bibitem{ritort} C.J. P\'erez Vicente and F. Ritort, J. Phys. A (1996)
submitted.

\end{references}
\end{document}